\documentclass[%
 reprint,
 superscriptaddress,
 amsmath,amssymb,
 aip,
 jap,
]{revtex4-1}

\usepackage{textcomp}
\usepackage[utf8]{inputenc}

\usepackage{graphicx}
\usepackage{dcolumn}
\usepackage{bm}
\usepackage[hidelinks]{hyperref}
\usepackage[version=4]{mhchem}
\usepackage{xcolor} 
\usepackage{longtable}

\newcommand\T{\rule{0pt}{2.6ex}}       
\newcommand\B{\rule[-1.2ex]{0pt}{0pt}} 

\begin{document}

\title{Towards unification of perovskite stability and photovoltaic performance
assessment}

\author{Bernard Wenger}
\email{bernard.wenger@physics.ox.ac.uk}
\affiliation{Clarendon Laboratory, Department of Physics, University of Oxford, Oxford OX1 3PU, United Kingdom}

\author{Henry J. Snaith}
\email{henry.snaith@physics.ox.ac.uk}
\affiliation{Clarendon Laboratory, Department of Physics, University of Oxford, Oxford OX1 3PU, United Kingdom}

\author{Isabel H. Sörensen}
\affiliation{accelopment AG, Seefeldstrasse 198, 8008 Zürich, Switzerland}

\author{Johannes Ripperger}
\affiliation{accelopment AG, Seefeldstrasse 198, 8008 Zürich, Switzerland}

\author{Samrana Kazim}
\affiliation{BCMaterials-Basque Center for Materials, Applications and Nanostructures, Martina Casiano, UPV/EHU Science Park, Barrio Sarriena s/n, Leioa 48940, Spain}

\author{Shahzada Ahmad}
\affiliation{BCMaterials-Basque Center for Materials, Applications and Nanostructures, Martina Casiano, UPV/EHU Science Park, Barrio Sarriena s/n, Leioa 48940, Spain}

\author{Edgar R. Nandayapa}
\affiliation{Fraunhofer Institute of Applied Polymer Research, Geiselbergstr. 69, 14476 Potsdam, Germany}

\author{Christine Boeffel}
\affiliation{Fraunhofer Institute of Applied Polymer Research, Geiselbergstr. 69, 14476 Potsdam, Germany}

\author{Silvia Colodrero}
\affiliation{Leitat Technological Center, Carrer de la Innovació 2, Barcelona, Spain}

\author{Miguel Anaya}
\affiliation{Cavendish Laboratory, Department of Physics, University of Cambridge, Cambridge CB3 0HE, United Kingdom}

\author{Samuel D. Stranks}
\affiliation{Cavendish Laboratory, Department of Physics, University of Cambridge, Cambridge CB3 0HE, United Kingdom}

\author{Iván Mora-Seró}
\affiliation{Institute of Advanced Materials (INAM), University Jaume I, Avenida de Vicent Sos Baynat, s/n, 12071 Castelló de la Plana, Spain}

\author{Terry Chien-Jen Yang}
\affiliation{École Polytechnique Fédérale de Lausanne, (EPFL), Institute of Microengineering (IMT), Photovoltaics and Thin-Film Electronics, Laboratory (PV-Lab), Rue de la Maladière 71b, 2002 Neuchâtel, Switzerland}

\author{Matthias Bräuninger}
\affiliation{École Polytechnique Fédérale de Lausanne, (EPFL), Institute of Microengineering (IMT), Photovoltaics and Thin-Film Electronics, Laboratory (PV-Lab), Rue de la Maladière 71b, 2002 Neuchâtel, Switzerland}

\author{Thorsten Rissom}
\affiliation{Oxford PV, Unit 7-8 Oxford Industrial Park, Mead Road, Oxford OX5 1QU, United Kingdom}

\author{Tom Aernouts}
\affiliation{Thin-Film Photovoltaics group, IMEC, Thor Park 8320, B-3600 Genk, Belgium}

\author{Maria Hadjipanayi}
\affiliation{FOSS Research Centre for Sustainable Energy, Photovoltaic Technology Laboratory, University of Cyprus, P.O. Box 20537, 1678, Nicosia, Cyprus}

\author{Vasiliki Paraskeva}
\affiliation{FOSS Research Centre for Sustainable Energy, Photovoltaic Technology Laboratory, University of Cyprus, P.O. Box 20537, 1678, Nicosia, Cyprus}

\author{George E. Georghiou}
\affiliation{FOSS Research Centre for Sustainable Energy, Photovoltaic Technology Laboratory, University of Cyprus, P.O. Box 20537, 1678, Nicosia, Cyprus}

\author{Alison B. Walker}
\affiliation{Department of Physics, University of Bath, Bath BA2 7AY, United Kingdom}

\author{Arnaud Walter}
\affiliation{CSEM PV-Center, rue Jaquet-Droz 1, 2002 Neuchâtel, Switzerland}

\author{Sylvain Nicolay}
\affiliation{CSEM PV-Center, rue Jaquet-Droz 1, 2002 Neuchâtel, Switzerland}

\date{\today}

\begin{abstract}
With the rapid progress of perovskite photovoltaics (PV), further challenges arise to meet meet the minimum standards required for commercial deployment. Along with the push towards higher efficiencies, we identify a need to improve the quality and uniformity of reported research data and to focus efforts upon understanding and overcoming failures during operation. In this perspective, as a large and representative consortium of researchers active in this field, we discuss which methods require special attention and issue a series of recommendations to improve research practices and reporting.
\end{abstract}

\maketitle

\section{Introduction}
Research into perovskite photovoltaics (PV) is currently advancing rapidly, making this technology a promising supplier of clean and affordable solar electricity for future low-carbon societies. So far, the focus has been primarily set on the pursuit of high power conversion  efficiencies (PCE), which reached above 25\% in single junction cells in 2019.\cite{NRELsBestResearch-CellEfficienciesChart} However, for this technology to meet the minimum standards required for commercial deployment, along with the push towards higher efficiencies, there is a need to improve the quality and uniformity of reported research data and to focus efforts upon understanding and overcoming failures during operation. Despite much discussion and suggestions, there remains a lack of community-wide agreement and adherence on the appropriate performance characterisation methods and, more significantly, upon the long-term simulated environmental stressing and stability assessment protocols. Therefore, as a large and representative consortium of researchers active in this field, we discuss which methods require special attention and issue a series of recommendations to improve research practices and reporting. In order to quantitatively compare improvements in material and solar cell stability between different laboratories, we believe it important that a small number of standardised stress tests are employed and reported upon. We believe that it is only by improving the quality and the reproducibility of research results that the field will reach the maturity required to deliver its full potential.

As a principle, the characterisation methods used for silicon and thin-film solar cells should apply to perovskites. However, these methods do not capture the peculiarities demonstrated for perovskite solar cells. For instance, this includes the frequent observation of hysteresis in current-voltage (\emph{JV}) curves, which depends on the bias direction, scan speed and pre-biasing conditions.\cite{Snaith2014} Moreover, there can be significant differences between the performance parameters extracted from such JV scans and those obtained from steady-state measurements, when the current density reaches a non-changing value at fixed voltage bias and illumination conditions. The distinctive behaviour of perovskite devices is often related to their slow response under fast varying electrical bias and illumination conditions, which has been largely discussed as resulting from ionic motion and charge accumulation in the active absorber layer.\cite{Snaith2014, Moriarty2017} In addition, the performance of the devices can sometimes be improved upon light soaking for up to several hours.\citep{Sahli2018} Although the community is converging towards a standardised measurement protocol in order to determine the steady-state power conversion efficiency of perovskite solar cells, this is not yet universally adopted.
Accelerated stress tests and stability assessment protocols are of much more variance than absolute performance measurements, and require significant consolidation. Specifically, a wide range of stressing conditions are applied to perovskite solar cells in order to assess the long term stability, which can range from cells being stored in an inert atmosphere at room temperature, to cells exposed to temperatures of 85\textdegree C and even 150\textdegree C, with additional aging factors such as light exposure and humidity. This massive divergence of degrees of stressing and final efficiency assessment within the research community, make it very difficult for other researchers or industry to make a fair judgement as to how stability improvements from one lab, relate to another, and to compare the merits of different approaches.

Here, we discuss which specific methods should be adopted, and under which circumstances. We address separately the assessment of stability and the measurement of the performance (efficiency), and propose some recommendations which, in our view, should be applied broadly amongst the research community. For the evaluation of stability, we recommend to follow the existing international standards and propose to adopt additional tests to address degradation modes more specific to perovskites. In the first place, light-induced degradation tests at specific elevated temperatures should be performed. Next, we propose the adoption of a select number of rigorous stability protocols, which is important for comparing stability advances across the community. Finally, we reaffirm the recommendation for ubiquitous use of steady-state figures of merit for determination and reporting of photovoltaic performance, as well as presenting the spectral response of the devices.

\section{Stability}

Long term operational stability and toxicity are often invoked as the major obstacles on the route to the commercialisation of perovskite PV. In this perspective, we do not specifically address the environmental impact of this technology but concentrate on the assessment of the stability of perovskite solar cells and modules. To evaluate stability, standard qualification tests for solar modules have been developed by international committees in order to measure the reliability of commercial PV products. In particular, the International Electrotechnical Commission IEC 61215 standard provides a series of fail/pass tests suitable for Si and thin-film technologies. Perovskite solar modules, as well as multi-junction devices incorporating perovskites, should therefore be subjected to the same tests. However, in practice these tests only provide a set of minimum requirements while current industrial specifications for stability go well beyond this standard. 

In an R\&D perspective, it is important to differentiate between qualification tests, such as those mentioned above and stability tests which objective is to identify failure modes of solar cells and modules, and develop routes to circumvent them. Due to the low stability of early perovskite devices and peculiar behaviour, such as for example the recovery of some devices during dark storage in ``day-night'' cycling simulations,\cite{Domanski2017} a number of different stability tests have been proposed in the literature, which has consequently led to some confusion. For example, prolonged illumination with a light source which lacks a field-relevant UV component or the use of inert atmosphere instead of encapsulation can prevent the identification of important degradation modes.\cite{Snaith2018} Therefore, along with other authors,\cite{Snaith2018, Christians2018} we point at the need for developing consistent stability tests in order to improve inter-laboratory comparisons and provide recommendations for future test standards. A significant attempt towards this objective was recently reported by a wide consortium of researchers, proposing a complete set of tests based on ISOS procedures.\cite{Khenkin2020} Furthermore, it would be very valuable to be able to appreciate objectively how a stability improvement reported by one lab compares with a stability improvement reported by another. To that end, there is a requirement for a small number of rigorous stress tests to be accepted as the standard for the research community, until further knowledge is known about yet-to-be-discovered failure modes. 

In line with the above, we recommend following the tests defined by the IEC 61215 standard in a preliminary stage (see for example Holzhey et al.\cite{Holzhey2018} for an illustration of this standard). We emphasise that the IEC 61215 is not the ``gold-standard'' in industrial PV stress testing, but is simply a ``stage-gate'' through which all modules must pass. PV manufacturers perform internal stress tests well beyond the IEC tests (for longer time, higher temperatures and more cycles). Therefore, if the perovskite research community performs tests which are less stringent than those of IEC 61215, this will not convince the existing PV industry that perovskite PV cells are stable enough to be considered for a potential future technology, and may even do more harm to the reputation of the technology than good. We also believe that even for less strenuous applications, such as indoor power, the IEC tests would have to be met in order to convince potential customers that short-term failure (\textless{} 1 year) will not be a problem. In Table \ref{tab:protocol}, we list a selection of stability tests that should reveal the most important degradation modes, as discussed previously by Snaith and Hacke.\cite{Snaith2018} The conditions for tests 2-4 are equivalent to the corresponding IEC 61215 tests. Test 1, is a relaxed version of test 4 that does not require environmental chambers, and thus is more easily applicable in research laboratories.

\begin{table}
\caption{\label{tab:protocol}Recommended conditions on device characterisation for
perovskite PV devices (based on the CHEOPS protocol)\cite{cheopsProtocol}}
\begin{ruledtabular}
\begin{tabular}{l p{6.5cm}}
\textbf{Category} & \textbf{Recommendation}\\ \hline \T
\emph{Irradiation} & Steady-state solar simulator with AM1.5G spectrum \B \\
\emph{Stability} & \B \\
1 & Stress devices at 85\textdegree C in the dark in inert atmosphere or air (encapsulated devices) \B \\
2 & Thermal cycling between -40\textdegree C and 85\textdegree C for 200 cycles \B \\ 
3 & Light soaking with simulated solar spectrum at 60°C or 85\textdegree C for 1000h at maximum power point (\emph{MPP})\B \\ 
4 & Damp-heat test at 85\% relative humidity and 85\textdegree C for  1000h \B \\ 
5 & Reverse bias cells to give a dark current density equivalent to  \emph{J\textsubscript{sc}} or \emph{J\textsubscript{mpp}} for 1 hour \B \\ 
\multicolumn{2}{l}{\emph{Cell performance}}  \B \\ 
1 & Dark \emph{JV} scans in reverse (\emph{V\textsubscript{oc}} to  \emph{J\textsubscript{sc}}) and forward (\emph{J\textsubscript{sc}} to \emph{V\textsubscript{oc}}) directions \B \\ 
2 & Light \emph{JV} scans at 1-sun illumination in reverse and forward  \B \\ 
3 & \emph{MPP} tracking (or continuous current measurement at fixed \emph{V\textsubscript{mpp}}) until steady-state (unchanging value) is reached. Typically, between 1 to 5 minutes. \B \\ 
4 & External quantum efficiency measurements \B \\
\end{tabular}
\end{ruledtabular}
\end{table}

In addition to the existing standards, stability tests specifically designed for perovskites solar cells in order to probe known failure modes are needed. We note that the standards for PV are occasionally updated with new tests, but this can take several years to decades to be implemented, as was the case for potential induced degradation (PID). The following is an outline of such ``beyond IEC'' tests, which we have identified from our experience, and points out where there is an evident need for further research and efforts of the PV community to drive the creation of such standards for measurements. 

Firstly, light soaking has been identified as a major stress factor for perovskite solar cells,\cite{Snaith2018} and should be investigated with high priority. For instance, photons of high energy (e.g. in the UV range) can break chemicals bonds in organic compounds contained in charge transport layers, encapsulants and perovskites themselves. Furthermore, ion and defect migration is accelerated under illumination, and thus all related degradation mechanisms are aggravated.\cite{Kim2018} Ideally, tests should be conducted with the devices under load (\emph{V\textsubscript{MPP}} or \emph{MPP} tracking) so as to ensure ion distributions and current densities relevant to operational conditions. The light spectrum should be close to the solar spectrum, while containing a UV component. In addition to the chemical bond breaking mentioned above, UV light can induce instabilities in metal oxides, such as TiO\textsubscript{2} and ZnO, often used as n-type charge extraction layers. This has been explained by the formation of electron traps under UV light, leading to reduced charge mobility, and also from photocatalytic oxidation in the case of titania.\cite{Leijtens2013} In practice, these effects can be mitigated by the use of an additional UV filter. However, if a UV filter is required, it should be included in these tests, \emph{i.e.} the light source should still contain the UV component and the UV filter should be used in conjunction. The efficiency with and without the UV filter should also be reported.

More recently, with improvements in the general stability of the device architectures, materials choice and perovskite compositions, the specific photochemistry of the metal halide perovskites has begun to reveal itself as a key remaining degradation factor, with the photo-generation of I\textsubscript{2} being the primary degradation path. This process is strongly accelerated by increased temperature. The light soaking test should therefore be conducted under moderate thermal stress (60 or 85\textdegree C) to accelerate degradation. Higher elevated temperatures would be advantageous for exploring thermally activated degradation mechanisms and performing more accelerated stress tests.

Additionally, we propose a specific test suitable for testing encapsulation and packaging techniques, based on the perovskite methylammonium lead iodide (MAPbI\textsubscript{3}), which is known to undergo relatively fast degradation when exposed to a combination of light, oxygen and moisture. Since most research laboratories have a standard procedure for fabricating thin films of MAPbI\textsubscript{3}, this test will be easy to put in practice. What we call the ``MAPI test'' consists in the fabrication of a sample containing a thin film of MAPbI\textsubscript{3} deposited via any method available in the research groups upon glass or a substrate of choice, and subsequently encapsulated. This sample can then be exposed to various stress factors (heat, humidity, light, thermal cycling, etc.) and characterised via visual inspection, or UV-Vis light absorption measurements, since the degradation products of MAPbI\textsubscript{3} always contain PbI\textsubscript{2}, which is easily identified by its yellow colour. This approach provides an efficient way to assess the efficacy of new encapsulation methodologies and is analogous to the calcium test\cite{nisato2001evaluating} used in organic PV and light emitting diodes to investigate barrier materials. In the calcium test, a thin calcium film is deposited in inert conditions and encapsulated. Since calcium is extremely reactive to water, forming transparent and insulating Ca(OH)\textsubscript{2}, water ingress can easily be quantified using optical transmission or electrical resistance measurements.\cite{Paetzold2003} 

Next, the effect of mechanical stress has been scarcely investigated and may become important, considering the potential for perovskites in flexible and lightweight modules for portable and building-integrated applications. This could for example be investigated with repeated bending tests. We note that flexibility tests will appear in the forthcoming revision of the IEC 61215-2 standard (expected May 2020). Finally, damage due to reverse biasing should be investigated in more depth. If a module contains faulty or shadowed cells, the operating current can exceed the short-circuit current (\emph{I\textsubscript{sc}}) of the faulty or shadowed cell, and force the affected cell into reverse bias.\cite{Bowring2018} In these conditions the cells dissipate power which can lead to heat damages. As this is a damage mode specific to modules, it has not frequently been assessed for perovskite solar cells.\cite{Bowring2018} However, as the technology matures towards commercialisation, it should be considered without delay. The ``hot-spot'' test in IEC 61215 requires a complete module, with individual cells mechanically shadowed while subject to 1 sun illumination under short-circuit conditions for between 1 to 5 hours. For small laboratory cells, the worst case scenario can be simulated by driving a reverse bias current density, with a magnitude of \emph{J\textsubscript{sc}}, through the cell. Therefore, we have added this test to our recommendations in Table \ref{tab:protocol} (test 5).

In summary, we recommend in the first instance to apply the test procedures described by the IEC 61215 standard without adjustments. Additionally, we advise to carry out tests specific to the identified degradation modes in perovskite solar cells and modules. These include (i) light soaking at 60\textdegree C or 85\textdegree C for 1000+ hours (with UV components), (ii) bending tests for flexible devices, (iii) reverse bias tests, and finally (iv) the ``MAPbI\textsubscript{3} test'' as a facile method to investigate encapsulation and packaging solutions (Table \ref{tab:protocol}). Finally, we emphasize that the stability tests conditions must be adequately reported, and propose a check-list aiming at improving the comparison between different studies (Table \ref{tab:check-list}).

\begin{table}
\caption{\label{tab:check-list}Stability measurements check-list}
\begin{ruledtabular}
\begin{tabular}{p{0.3\columnwidth} p{0.6\columnwidth}}
\textbf{Parameters} & \textbf{Information to report} \T \B \\ 
\hline \T
Light source & Yes/No + Spectrum (with or without UV component), intensity \B  \\
Number of devices tested & Number  \B \\
Statistical analysis of the device & Yes/No, description \B \\
Environment & Atmosphere (N\textsubscript{2}, air, \ldots), humidity, temperature \B \\
Device status & \emph{MPP} tracking / static load / Open-circuit, short-circuit\cite{Chen2019a} \B \\
Encapsulation & Yes/No, method \B \\
Performance metrics & Absolute/relative \emph{PCE}/\emph{FF}/\emph{J\textsubscript{sc}}/\emph{V\textsubscript{oc}}/\emph{MPP} or Steady-state power output. \emph{MPP} or \emph{SPO} mandatory for comparing stability trends \B \\
Pre-conditioning & Yes/No, method \B \\
\end{tabular}
\end{ruledtabular}
\end{table}

\section{JV curves and steady-state power efficiencies}

Along with publishers\cite{natmater} and other researchers,\cite{Christians2015,Wang2019,Zimmermann2016} we emphasise the importance to comprehensively report measurement conditions and protocols used to obtain the JV curves. This will improve reproducibility among research groups and also trust in published results. Of particular importance to perovskite devices, reports should include scan rates, scan directions, pre-biasing, illumination conditions (including light soaking), device history and environment (temperature, atmosphere) as discussed in Table 3.

The phenomenon of anomalous hysteresis\cite{Snaith2014} in the current-voltage (\emph{JV}) curves has complicated the interpretation of the device performance in terms of the standard metrics(\emph{J\textsubscript{sc}}, \emph{V\textsubscript{oc}},\emph{P\textsubscript{max}}, \emph{FF}, \emph{J\textsubscript{MPP}},\emph{V\textsubscript{MPP}}).\textsuperscript{20} In some cases, the maximum power conversion efficiency obtained from a fast scan from forward bias back towards short-circuit can be several percent higher than the scan from short-circuit back towards open-circuit. This can be explained by the formation of transient injection barriers resulting from an unfavourable distribution of ions across the perovskite film during the fast forward scan, and the subtle interplay between ions and charge recombination centres in the perovskite absorber layer.\cite{VanReenen2015} Moreover, it is often observed that the maximum power output increases when the device is maintained at a bias close to \emph{V\textsubscript{MPP}}. Therefore, standard JV scanning protocols optimised for silicon PV usually can't be applied apply to perovskite solar cells. As a consequence, no uniform characterisation protocol has been adopted by the community, despite informal efforts to suggest best practices.\cite{Snaith2014,Unger2014} To further illustrate this difficulty in instituting a standardised procedure, we note that the different accredited independent solar cell certification labs do not share a common protocol. For example, NREL have developed a protocol to overcome the artefacts related to hysteresis where, instead of tracking the maximum power point (\emph{MPP}), the device's photocurrent is measured at a sub-set of fixed bias voltages close to the \emph{JV}-determined \emph{V\textsubscript{max}}.\cite{Moriarty2017} The slope of the photocurrent over time is repeatably evaluated at a certain fixed voltage (using a sliding time window) until it reaches a minimum threshold value (\emph{e.g.} 0.1\% of the approximate \emph{J\textsubscript{sc}}). The ``asymptotic'' photocurrent limit is then extrapolated via fitting with an exponential curve and the maximum power is interpolated from the voltage and asymptotic current pairs. Although we encourage research groups to implement such methods, this protocol is relatively complicated to put in practice and time consuming. Therefore, it is not applicable for day-to-day characterisation, which require faster and less sophisticated methods.

To overcome these challenges and ambiguities, we urge research groups to provide steady-state power conversion efficiencies, which are independent of the scan conditions. Indeed, it is commonly observed that the steady-state power output can be significantly different from the maximum power point calculated from a \emph{JV} scan curve due to the hysteretic character of such measurements. Ideally, steady-state power conversion efficiencies are obtained using an algorithm which will adjust the bias voltage so as to maximise the power output. \emph{MPP} tracking algorithms are used across a variety of PV technologies for the optimisation of the energy yield of modules under continuous real-life operation, as illumination and temperature conditions vary during the day. However, for slowly responding devices, such as perovskite solar cells, \emph{MPP} algorithms are also useful to track the maximum efficiency under standard test conditions compensating for ionic motion/charge accumulation phenomena. A variety of \emph{MPP} tracking algorithms have been developed. Typically, they apply small steps in the bias voltage, for instance starting close to the \emph{V\textsubscript{MPP}} determined from a \emph{JV} scanning curve, and then track the product of voltage and current. If the product is greater than the last step, another step is taken in the same direction, if the product is less than the last step, then a step is taken in the reverse direction, until the programme is oscillating between two or three voltage points. The magnitude of the voltage steps and time interval between each step depend on the specific response characteristics of the device. If the voltage steps are too big, or if the time left for the system to reach a new steady-state is too short, the algorithm can easily end up in a non-converging loop.

The advantage of such \emph{MPP} tracking is that the absolute maximum efficiency will be found, if the algorithm is effective. However, as a close approximation (always an underestimation), \emph{MPP} tracking methods can be replaced by fixed voltage steady-state current measurements, where the \emph{MPP} voltage (\emph{V\textsubscript{MPP}}) is determined from the \emph{JV} scan curve. The latter method is straightforward to implement and independent of an \emph{MPP} tracking algorithm. The duration of such steady-state methods depends on the device being tested. This is influenced for example by the ion diffusion coefficient or a potential performance improvement (or decay) upon light soaking. It should lead to a value which is non-changing in the short term, which is typically attained within 60 seconds for state-of-the-art perovskite solar cells.\cite{McMeekin2019}

Other measurement factors affecting the power conversion efficiency include temperature, 2- or 4-wire connections, repeated \emph{JV} scans, light soaking and atmospheric exposure, as well as the device history to electrical bias or light.

 In Table \ref{tab:protocol} we show recommended measurement conditions, based on a protocol reported previously (the CHEOPS protocol).\cite{cheopsProtocol} We are reluctant to provide a specific protocol for \emph{JV} scan rate, dwell time and range, and steady-state measurement time, since some parameters depend on the device response time, such as scan rate or direction, and need to be adjusted for a given set of devices. For significant results and record devices, we still recommend to obtain certifications from accredited laboratories. However, for most scientific reports and routine measurements, the protocol described in Table \ref{tab:protocol} is relatively easy to apply and will produce more reproducible and consistent metrics. As a general rule for \emph{JV} curves, measurement standards such as use of a solar simulator with homogeneous spatial intensity, spectral output and temporal stability (\emph{e.g.} AAA-class), and proper light source calibration and mismatch factor estimation should be a prerequisite.\cite{Snaith2012a} As previously reported,\cite{Snaith2012} the systematic error associated with inaccurate spectral mismatch factors and solar simulator calibration can largely exceed the random error which is usually reported from the standard deviation over a statistically significant number of devices.

In Table \ref{tab:params} we describe how testing conditions affect the PV metrics of the device and issue some recommendations. Importantly, such factors must be specified in the reports.

\begin{table*}
\caption{\label{tab:params}Measurement parameters influencing the determination of solar cell performance}
\begin{ruledtabular}
\begin{tabular}{p{0.1\textwidth} p{0.18\textwidth} p{0.32\textwidth} p{0.4\textwidth}}
Parameter & \textbf{Metrics predominantly affected} & \textbf{Recommendation} & Comments \T\B \\ \hline \T
Scan rate & \emph{V\textsubscript{MPP}}, \emph{J\textsubscript{MPP}}, \emph{V\textsubscript{oc}, J\textsubscript{­sc}, FF} & Adapt to device and always provide steady-state metrics for the same device. & Fast scan rates may not allow for the stabilisation of the distribution of ions throughout the perovskite films. This usually is illustrated by hysteresis between the \emph{JV} curves scanned in the forwards and reverse direction, and leads to errors in the estimation of all parameters.\cite{Snaith2014} \B \\
Scan direction & \emph{V\textsubscript{MPP}}, \emph{J\textsubscript{MPP}}, \emph{V\textsubscript{oc}, J\textsubscript{SC},} & Measure \emph{JV} curves in both directions (increasing and receding bias) & Hysteresis is easily identified when comparing both scans.\cite{Snaith2014} \B \\
Steady-state PCE & \emph{V\textsubscript{MPP}}, \emph{J\textsubscript{MPP}} & Measure photocurrent at fixed voltage (near \emph{V\textsubscript{MPP}}) or use \emph{MPP} tracking until a steady-state value is observed. & The fixed voltage can be determined from the \emph{JV} curve. Robust \emph{MPP} tracking algorithms must be used to prevent oscillations around the \emph{MPP}. \B \\
Voltage Pre-bias & \emph{V\textsubscript{MPP}}, \emph{J\textsubscript{MPP}}, \emph{FF}\textsubscript{,} \emph{V\textsubscript{oc}} & No pre-biasing for forward scan. Stabilisation at \emph{V\textsubscript{oc}} for backward scan. & With these recommendations, the device is pre-conditioned in the corresponding steady-state situation. Pre-biasing at voltage beyond \emph{V\textsubscript{oc}} can lead to artificially high \emph{FF}, since ion accumulation in these conditions can promote charge extraction.\cite{VanReenen2015} \B \\
Light-soaking & \emph{V\textsubscript{MPP}}, \emph{J\textsubscript{MPP}}, \emph{V\textsubscript{oc}, J\textsubscript{­sc}, FF} & If light soaking is performed before the measurement, this must be specified. & Prolonged illumination can lead both to increase performance (\emph{e.g.} via passivation) or degradation.\cite{Snaith2018,Nie2016} \B \\
Temperature & & Use a temperature-controlled stage. Standard temperature is 25\textdegree C. & Solar cell efficiency usually decreases with increasing temperature.\citep{Dupre2015} \B \\
Wiring & Series resistance (\emph{R\textsubscript{s}}) & Use 4-terminal sensing & 4-terminal wiring eliminates lead and contact resistances from the measurements. These effects are predominantly observed for large area devices due to the increase in photocurrent. \B \\
Masking & \emph{V\textsubscript{MPP}}, \emph{J\textsubscript{MPP}}, \emph{V\textsubscript{oc}, J\textsubscript{­sc}, FF} & Use a non-reflective mask with same shape, but fractionally smaller than the electronic active area & Without masking, excess light can diffuse towards the active area leading to overestimation of current density.\cite{Wang2019,Snaith2012a} Moreover, with mask apertures much smaller than area defined by the overlapping electrodes, \emph{V\textsubscript{oc}} is underestimated and \emph{FF} overestimated.\cite{Kiermasch2018a} \B \\
Repeated \emph{JV} scans & & Same conditions as initial \emph{JV} scan (scan rate, scan direction, pre-biasing). & Repeating \emph{JV} scans, provides a facile way to identify the presence of short-term degradation or improvements due to light-soaking. \B \\
Atmosphere & & State whether encapsulated or non-encapsulated, and atmosphere in which measurement is performed. & Oxygen and humidity have been shown to induce both passivation and/or degradation.\cite{Bryant2015,Brenes2018} \B \\
\end{tabular}
\end{ruledtabular}
\end{table*}

\section{Spectral response}

The external quantum efficiency (EQE) of a PV device corresponds to the incident photon-to-electron conversion efficiency measured under short-circuit current conditions as a function of wavelength. This measurement can provide insight into the processes which limit the efficiency of photovoltaic devices, such as charge collection efficiency, parasitic absorption and film thickness limitations, and as such should be investigated by researchers to understand and improve their devices. Moreover, the determination of the spectral response for a given type of solar cells is a requisite for the calculation of the mismatch factor correction required for the accurate calibration of the solar simulator.\cite{Snaith2012} However, we would like to emphasise the importance to present these measurements in every report containing performance of devices, since integration of the EQE spectrum over the AM1.5G solar spectrum provides a complementary measurement of the short-circuit photocurrent density generated by the device. This is important since the calibration of a solar simulator can be a complicated and sensitive task,\cite{Snaith2012} which easily leads to systematic errors. In this context, the integrated current from an EQE spectrum provides an independent estimate of \emph{J\textsubscript{sc}}. For multi-junction cells, EQE spectra are absolutely essential due to the high sensitivity of the measurement to the calibration of the solar simulator. As these are subtler and more complicated characterisation methods, we address them separately for multi-junction devices below.

We are aware of the fact that the uniformity of the light source in an EQE setup is usually not as well defined as it is in a solar simulator. The standard practice for measuring the EQE is for the monochromatic light source to be focused to a smaller area than the active area of the PV cell, and the calibration reference cell. By this means, the total monochromatic light flux is incident upon the test and reference cells, and accuracy errors associated with active area definition by opaque optical masks and light source uniformity are obviated. However, some lab-based EQE setups have illumination areas larger than the active area of the solar cell, and rely on optical masking to define the active area of the solar cell. Small errors in mask alignment can lead to substantial differences between the test and reference devices, especially when testing small-area cells (\textless{} 1 cm\textsuperscript{2}) as often encountered in perovskite research. We recommend against this practice, and advise ensuring that the monochromatic light source is focused to a smaller area than both the test and reference cells.

Some research groups use a white light bias illumination to bring the samples closer to the carrier injection levels experienced in operational conditions, and the small area monochromatic light can be considered as a minor perturbation. This has the advantage that the EQE is being estimated in conditions similar to what the solar cell experiences under sun light. This is particularly important if the cells exhibit a non-linearity in photocurrent density with light intensity.

We also note that perovskite solar cells sometimes show a reduced steady-state power output with respect to the efficiency determined from the \emph{JV} curve. This is usually observed for devices with slow responses. In such cases, they may exhibit a lower steady-state \emph{J\textsubscript{sc}}, in comparison to the \emph{JV}-determined \emph{J\textsubscript{sc}}. Since the EQE measurement is a pseudo-steady-state measurement, we expect this to be closer matched with the steady-state \emph{J\textsubscript{sc}} (in comparison with the \emph{J\textsubscript{sc}}-determined from a fast \emph{JV} measurement). We suggest measuring the steady-state \emph{J\textsubscript{sc}} under the AM1.5G solar simulator, in order to accurately compare with the EQE-determined \emph{J\textsubscript{sc}}.

\section{Tandem and multi-junction device characterisation}

One of the most exciting prospects of perovskite PV technology is its potential for tandem solar cells in combination with crystalline silicon or other thin film PV materials.\cite{Han2018a} As of today, this approach appears best suited for a large commercial deployment of perovskites in PV and is currently pursued by several consortia and industrial players.\cite{Snaith2018a} Furthermore, the ability of perovskite absorbers to be designed with tuneable band gaps opens the possibility to build ``perovskite-only'' multi-junction devices. For JV curves and steady-state efficiencies the spectral mismatch between the calibrated reference diode and the device under investigation can be complicated to determine accurately, as we discuss below. In addition, the measurement of EQE of the independent sub-cells is challenging, especially when the absorption spectra of the active layers overlap. These issues as well as best measurement practices have been reported previously for III-V multi-junction devices\cite{Meusel2003,Meusel2002,ASTM-E2236} as well as for organic PV cells.\cite{Timmreck2015} In principle, the methods developed for these technologies are applicable and should be accurately followed, considering the particularities of perovskite solar cells when acquiring JV curves for the calculation of the mismatch factors.

\begin{figure}
\includegraphics[width=\columnwidth]{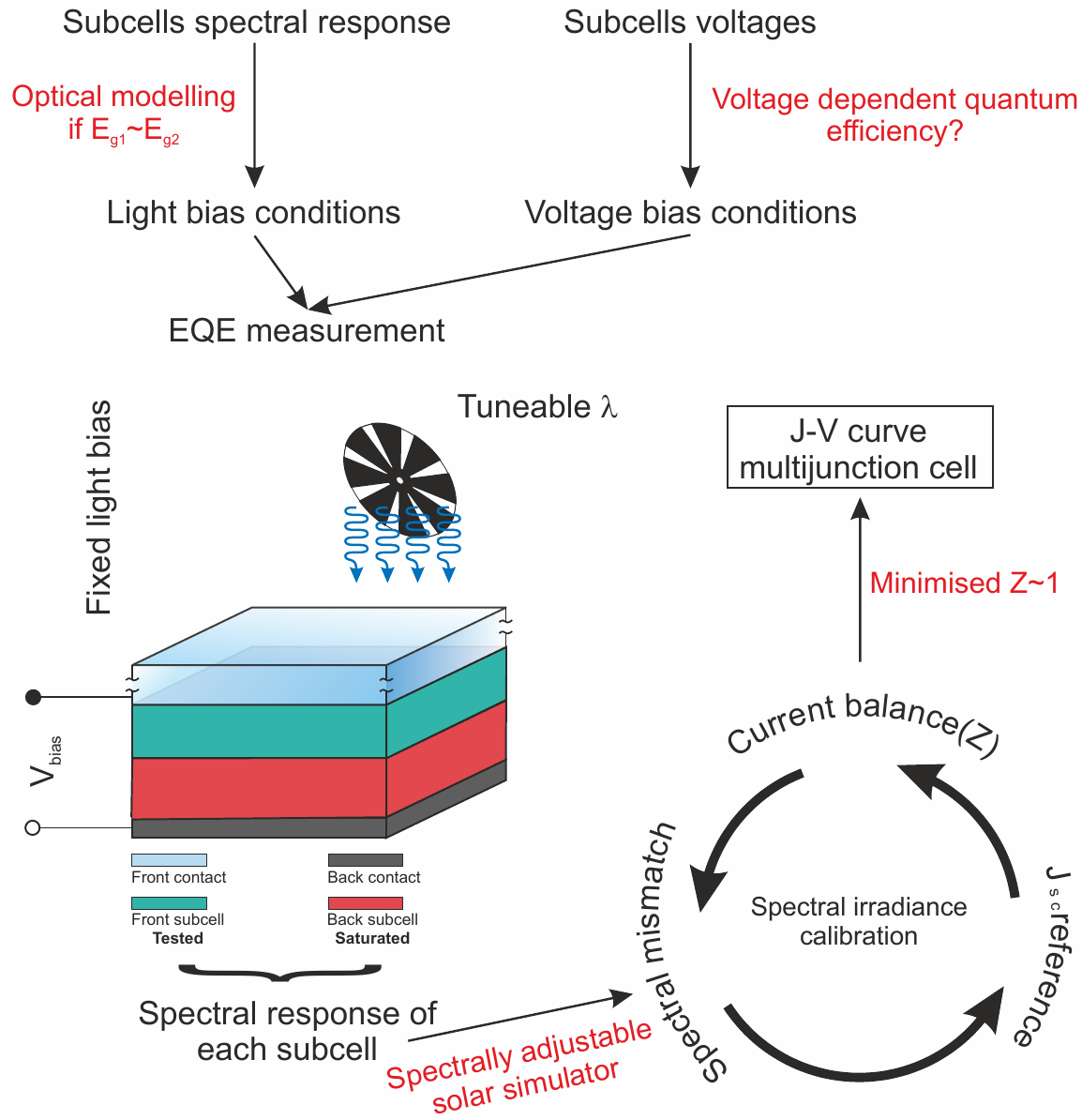}
\caption{\label{fig1}Schematic description of the procedure to characterise the power conversion efficiency of 2-terminal multi-junction solar cells.}
\end{figure}

Hereafter, we briefly describe the recommended procedure to obtain reliable performance characterisation of 2-terminal multi-junction solar cells: the global scheme is depicted in Figure \ref{fig1}. In order to determine the performance of the device under standard test conditions (i.e. AM1.5G, 100 mW/cm\textsuperscript{2}), the spectrum of the solar simulator has to be adjusted so that each sub-cell generates the same photocurrent as under the AM1.5G reference spectrum. This can be achieved with a dual or multi-zone solar simulator, with each zone chosen to match the spectral response of each sub-cell. Each sub-spectrum can then be adjusted mathematically with a spectral mismatch correction which requires the knowledge of the solar simulator spectrum and the spectral response of each sub-cell. Therefore, the first stage consists in the accurate measurement of the spectral response (or external quantum efficiency, EQE) for each sub-cell. In practice, this can be achieved for a given sub-cell by saturating all other sub-cells with monochromatic bias illumination so that the photocurrent of the full stack is limited by the response of the sub-cell under test. In case of overlap of the sub-cells' spectral range, as commonly observed for organic solar cells, optical modelling is required to determine their relative contributions to the photocurrent. In addition, voltage biasing needs to be applied so that the sub-cell under test runs near short-circuit conditions, and also to prevent artefacts appearing in the case of low shunt resistance or low reverse breakdown voltage.\cite{Meusel2003} Such artefacts can be identified by performing a dark spectral response (i.e. when no bias light is applied) measurement on the multijunction device.\cite{Pravettoni2011} Once the spectral responses have been obtained, a mismatch factor can be calculated for each sub-cell. Then in order to generate the required photocurrent in each sub-cell the spectrum of the solar simulator must be adjusted and tested by measuring the \emph{J\textsubscript{sc}} of a reference solar cell under the new spectral conditions. New mismatch factors are then calculated and this process must be run iteratively until a consistent photocurrent is obtained for the reference solar cell. This process is only achievable by using a multi-source solar simulator, dual-source for tandem cells, in which the spectra of the light sources are fixed but the intensities can be adjusted independently.\cite{Meusel2002} If the overall spectrum of the light source is fixed, then although the mismatch factor can be applied to an accurate estimate of the \emph{J\textsubscript{sc}}, it is highly likely that the current density will be mismatched between the multi-junctions, and the \emph{FF} of the multi-junction solar cell will be overestimated, in comparison to what would be obtained under true AM1.5G.

\section{Conclusions}

In this perspective, we have discussed the lack of community-wide agreement upon simulated and accelerated environmental stress test and performance characterisation methods which need to be applied for perovskite solar cells and modules. We issue recommendations on the stability test protocols and measurement procedures to adopt. For the assessment of stability, we recommend to follow the international standards developed for silicon and thin film photovoltaics in the first instance, as outlined in IEC 61215. However, in order to address degradation modes more specific to perovskites, we recommend additional tests. First and foremost, light-induced degradation should be addressed with extended (1000+ hours) light soaking tests at elevated temperature. We also point at other concerns, such as the failure of flexible cells due to mechanical stress, as well as reverse biasing in cells and modules. The issues mentioned in this perspective can only be addressed meticulously, through inclusive discussions with the participation of the entire perovskite PV community. We plead for the research community to report steady-state performance values and to comprehensively report the scanning conditions used to obtain JV curves. We also discuss the pitfalls associated with the measurement of EQE and summarised the requirements for the characterisation of multi-junction solar cells. We are confident however, that solutions are in sight and that collaboration is the answer to further advancement, since we see perovskite PV as having the potential to be the corner stone of a new clean and affordable energy system in the near future.

\acknowledgments
On 25-26 June 2018, CHEOPS (https://www.cheops-project.eu) took the initiative to invite all major EU-funded perovskite PV projects to the European Perovskite PV Days 2018 to discuss the topics of device characterization and stability, environmental impact and commercial challenges. This perspective results from the discussions held during this meeting and corresponds to a common view shared by all participants. The authors have received funding from the European Union's Horizon 2020 research \& innovation programme under grant agreements No. 763989 (APOLO), 706552 (APPEL), 653296 (CHEOPS), 764047 (ESPResSo), 797546 (FASTEST), 756962 (HYPERION), 764787 (MAESTRO), 724424 (No-LIMIT), 763977 (PerTPV), 747221 (POSITS) and 726360 (MOLEMAT). The research leading to these results has received funding from the European Union's Horizon 2020 research and innovation programme under grant agreement No. 653296 of the CHEOPS project and No. 763977 of the PerTPV project.

All authors participated to the discussions leading to the preparation
of the perspective (see Acknowledgments above). The redaction of the
manuscript was coordinated and executed by B.W. and H.J.S., and reviewed
by all authors. M.A. prepared the figure.

\bibliography{main}

\end{document}